\documentclass[sigconf]{acmart}

\AtBeginDocument{%
  }

\copyrightyear{2024}
\acmYear{2024}
\setcopyright{acmlicensed}
\acmConference[WSDM '24]{Proceedings of the 17th ACM International Conference on Web Search and Data Mining}{March 4--8, 2024}{Merida, Mexico}
\acmBooktitle{Proceedings of the 17th ACM International Conference on Web Search and Data Mining (WSDM '24), March 4--8, 2024, Merida, Mexico}
\acmPrice{15.00}
\acmDOI{10.1145/3616855.3635774}
\acmISBN{979-8-4007-0371-3/24/03}

\usepackage{url}
\usepackage{multirow}
\usepackage{multicol}
\usepackage{subfigure}
\usepackage{booktabs}
\usepackage{color}
\usepackage{makecell}
\usepackage{graphicx}
\usepackage{hyperref}

\usepackage{amsmath,amsfonts,bm}

\def\eqref#1{equation~\ref{#1}}

\def\1{\bm{1}}

\def\vj{{\bm{j}}}

\def\vp{{\bm{p}}}
\def\vq{{\bm{q}}}

\def\vs{{\bm{s}}}
\def\vt{{\bm{t}}}

\def\vw{{\bm{w}}}

\def\vy{{\bm{y}}}

\DeclareMathAlphabet{\mathsfit}{\encodingdefault}{\sfdefault}{m}{sl}
\SetMathAlphabet{\mathsfit}{bold}{\encodingdefault}{\sfdefault}{bx}{n}

\def\sP{{\mathbb{P}}}
\def\sQ{{\mathbb{Q}}}

\newcommand{\Ls}{\mathcal{L}}

\newcommand{\KL}{D_{\mathrm{KL}}}

\usepackage[noabbrev,capitalize]{cleveref}
\acmSubmissionID{170}

\begin{document}

\title{LEAD: Liberal Feature-based Distillation for Dense Retrieval}

\author{Hao Sun}
\affiliation{
  \institution{Peking University}
  \city{Beijing}
  \country{China}
}
\email{sunhao@stu.pku.edu.cn}

\author{Xiao Liu}
\affiliation{
  \institution{Microsoft Research Asia}
  \city{Beijing}
  \country{China}
}
\email{xiao.liu.msrasia@microsoft.com}

\author{Yeyun Gong}
\affiliation{
  \institution{Microsoft Research Asia}
  \city{Beijing}
  \country{China}
}
\email{yegong@microsoft.com}

\author{Anlei Dong}
\affiliation{
  \institution{Microsoft Corporation}
  \city{Mountain View}
  \state{CA}
  \country{USA}
}
\email{anlei.dong@microsoft.com}

\author{Jingwen Lu}
\affiliation{
  \institution{Microsoft Corporation}
  \city{Redmond}
  \state{WA}
  \country{USA}
}
\email{jingwen.lu@microsoft.com}

\author{Yan Zhang}
\affiliation{
  \institution{Peking University}
  \city{Beijing}
  \country{China}
}
\email{zhyzhy001@pku.edu.cn}

\author{Linjun Yang}
\affiliation{
  \institution{Microsoft Corporation}
  \city{Redmond}
  \state{WA}
  \country{USA}
}
\email{yang.linjun@microsoft.com}

\author{Rangan Majumder}
\affiliation{
  \institution{Microsoft Corporation}
  \city{Redmond}
  \state{WA}
  \country{USA}
}
\email{ranganm@microsoft.com}

\author{Nan Duan}
\affiliation{
  \institution{Microsoft Research Asia}
  \city{Beijing}
  \country{China}
}
\email{nanduan@microsoft.com}

\begin{abstract}
Knowledge distillation is often used to transfer knowledge from a strong teacher model to a relatively weak student model.
Traditional methods include response-based methods and feature-based methods. Response-based methods are widely used but suffer from lower upper limits of performance due to their ignorance of intermediate signals, while feature-based methods have constraints on vocabularies, tokenizers and model architectures.
In this paper, we propose a \textbf{l}iberal f\textbf{e}ature-b\textbf{a}sed \textbf{d}istillation method (\textbf{LEAD}). LEAD aligns the distribution between the intermediate layers of teacher model and student model, which is effective, extendable, portable and has no requirements on vocabularies, tokenizers, or model architectures.
Extensive experiments show the effectiveness of LEAD on widely-used benchmarks, including MS MARCO Passage Ranking, TREC 2019 DL Track, MS MARCO Document Ranking and TREC 2020 DL Track.
Our code is available in \url{https://github.com/microsoft/SimXNS/tree/main/LEAD}.
\end{abstract}


\maketitle

\section{Introduction}

Recently, with the advent of large pre-trained language models~\cite{devlin2018bert,brown2020language,smith2022using, sun2019ernie20}, dense retrieval \cite{DBLP:conf/emnlp/KarpukhinOMLWEC20,DBLP:conf/iclr/XiongXLTLBAO21,DBLP:journals/corr/abs-2210-11773,ni2021large} has become a popular topic for information retrieval, which aims to identify relevant contents from a large collection of passages.
Despite the great success of dense retrieval on benchmarks like MS MACRO~\cite{DBLP:conf/nips/NguyenRSGTMD16}, the inference efficiency is still left as a problem.
Therefore, small and efficient models are welcomed in the practical scenarios of dense retrieval.

Considering the trade-off between efficiency and effectiveness, researchers resort to knowledge distillation (KD) \cite{sanh2019distilbert,jiao2019tinybert, phuong2019towards, tang2020understanding}, which aims to transfer knowledge from a strong and large teacher model to a small yet efficient student model.
Popular distillation methods involve response-based methods and feature-based methods.
Response-based methods \cite{DBLP:journals/corr/HintonVD15, kim2018paraphrasing, mirzadeh2020improved} train the student model to predict the output of the teacher model, which is simple yet effective in different tasks.
For example, \cite{lin2022prod,DBLP:conf/sigir/ZengZV22,DBLP:journals/corr/abs-2205-09153} combine knowledge distillation with curriculum strategies to gradually improve the performance of the student model.
\cite{DBLP:conf/emnlp/RenQLZSWWW21,DBLP:conf/iclr/ZhangGS0DC22} propose on-the-fly distillation, which allows for the simultaneous optimization of both the teacher and student models.
However, compared with feature-based methods, response-based methods ignore the signals from the intermediate layers of teachers, which are proven to be important in deep and thin neural networks~\cite{DBLP:journals/corr/RomeroBKCGB14} and lead to lower upper limits than feature-based methods.

Feature-based methods\cite{DBLP:journals/corr/RomeroBKCGB14,DBLP:conf/iclr/ZagoruykoK17,DBLP:conf/iclr/ZagoruykoK17,passalis2018learning} mainly use the representations or attention maps of the intermediate layers as features, directly matching the features of the teacher model and the student model.
For example, \cite{DBLP:journals/corr/RomeroBKCGB14,xu2020feature,wang2020exclusivity,sun2019patient,haidar2021rail} employ direct matching of the internal representations between the teacher and the student.
\cite{DBLP:conf/iclr/ZagoruykoK17, passban2021alp, huang2017like,li2021virt} focus on enclosing the attention maps of the teacher and the student.
However, these methods suffer from two major weaknesses:
(1) In our preliminary study, we have observed that feature-based methods do not perform well in dense retrieval tasks, often yielding lower performance compared to response-based methods. This suggests that feature-based approaches may not be suitable for dense retrieval scenarios, highlighting the need for alternative methodologies.
(2) Matching features between the teacher model and student model requires them to share the same tokenizer-vocabulary set or have identical internal output sizes. This constraint can limit the flexibility in choosing the teacher model, as it must align with the requirements of the student model.

To overcome the limitations of  both response-based methods and feature-based methods, we propose a liberal feature-based distillation method (LEAD) for dense retrieval.
Specifically, we use the internal \texttt{[CLS]} embedding vectors to calculate the similarity distribution of the passages as features.
Then, we align the features between the teacher and the student, to accomplish the feature matching without tokenizer or architecture constraints.
We randomly select layers from the teacher model and student model to participate in the feature matching.
However, considering that the informativeness of selected teacher layers may be different, we propose a layer reweighting technique that adaptively assigns weights to the selected teacher layers based on its informativeness.
Besides, we jointly train the teacher model and student model to let them learn from each other.
The reason why we call our method \textbf{liberal} comes from two folds:
(1) LEAD supports various teacher architectures like dual encoder, ColBERT, cross encoder, and so on, only assuming that the teacher has a layer number no less than the student.
(2) LEAD uses the label distributions to calculate distillation loss, which is free of the constraints on vocabularies, tokenizers, or model architectures.

To evaluate the effectiveness of LEAD, we conduct experiments on several benchmarks, including MS MACRO Passage Ranking~\cite{DBLP:conf/nips/NguyenRSGTMD16}, TREC 2019 DL Track~\cite{DBLP:journals/corr/abs-2003-07820}, MS MARCO Document Ranking~\cite{DBLP:conf/nips/NguyenRSGTMD16} and TREC 2020 DL Track~\cite{DBLP:journals/corr/abs-2102-07662}.
The experimental results demonstrate that LEAD consistently outperforms traditional response-based and feature-based methods on these benchmarks, verifying its effectiveness in the context of dense retrieval.
Furthermore, we conduct an ablation study to evaluate the impact of our joint training and layer reweighting technique. The results confirm the effectiveness of these techniques, highlighting their contribution to the improved performance of LEAD.
Additionally, we perform a layer selection study to assess the effectiveness of our random layer selection strategy. 
The study demonstrates that the random layer selection strategy outperforms other layer selection strategies, further validating the efficacy of our approach.
Moreover, we observe two key advantages of LEAD:
(1) LEAD is extendable to broad-sense ``layers'', which include Transformer layers and linear layers that can be appended to models depending on the downstream tasks.
(2) LEAD is portable for the existing popular distillation pipelines for dense retrieval \cite{DBLP:conf/sigir/ZengZV22,lin2022prod}, which means it not only achieves superior performance by itself but also can achieve further improvements with other effective pipelines, such as curriculum learning~\cite{bengio2009curriculum,lin2022prod,DBLP:conf/sigir/ZengZV22,he2022curriculum} and data augmentation~\cite{shorten2019survey,DBLP:conf/emnlp/RenQLZSWWW21,bonifacio2022inpars}.

The contributions of this work are as follows:
\begin{itemize}
    \item We propose a novel knowledge distillation method LEAD, which uses similarity distribution of intermediate layers to distill knowledge from teacher model to student model. We use the \texttt{[CLS]} embedding vectors in the intermediate layer to calculate the similarity distribution, which accomplishes feature matching without the constraint on tokenizer or model architecture.
    \item We propose an effective layer aggregation method to let the student model learn from the most informative layers of the teacher model. We propose to jointly train teacher model and student model to further improve the performance of teacher model. 
    \item LEAD achieves superior performance compared with both response-based methods and feature-based methods. LEAD is extendable and supports appending multiple linear layers after teacher model and student model. LEAD is easy-to-plug-in in existing distillation pipelines and further improves the performance of these distillation pipelines.
\end{itemize} 

\section{Related Work}


\subsection{Dense Retrieval}
The goal of text retrieval task is to find relevant passages given a query.
There are two categories of methods in text retrieval task, which include sparse retrieval methods and dense retrieval methods.
Sparse retrieval methods aim to identify relevant passages based on text matching.
For example, BM25 is a bag-of-words retrieval function that ranks a set of passages based on the query terms appearing in each passage.
Though sparse retrieval has high efficiency, it cannot identify semantically relevant but text-independent query-passage pairs.

As the contrary, dense retrieval methods \cite{DBLP:conf/sigir/ZhanM0G0M21, lin2021batch, lin2022prod, zhou2022master} can identify semantically relevant but text-independent query-passage pairs.
The query and passage representations in dense retrieval models are usually obtained using dual encoders \cite{DBLP:conf/emnlp/KarpukhinOMLWEC20}.
\cite{DBLP:conf/sigir/KhattabZ20} pushes performance higher by introducing the late interaction while requiring higher computation and storage costs.
In spite of the model structure, recent work also focuses on utilizing training strategies to obtain better results, ranging from data-centric studies~\cite{DBLP:conf/acl/RenLQLZSWWW21, sciavolino2021simple} and negative sampling~\cite{gao2021your, DBLP:conf/iclr/XiongXLTLBAO21, DBLP:journals/corr/abs-2210-11773} to distillation~\cite{DBLP:journals/corr/abs-2010-02666, DBLP:conf/sigir/ZengZV22, DBLP:journals/corr/abs-2205-09153}.
\cite{DBLP:conf/acl/RenLQLZSWWW21} proposes a dense retrieval approach that incorporates both query-centric and passage-centric similarity relations.
\cite{DBLP:conf/sigir/HofstatterLYLH21} proposes an efficient topic-aware query and balanced margin sampling technique, for training dense retrieval models.
\cite{DBLP:conf/sigir/ZengZV22} proposes a curriculum learning-based optimization framework, for enhancing dense retrieval models through knowledge distillation.
\cite{DBLP:journals/corr/abs-2205-09153} introduces on-the-fly distillation and cascade distillation processes for cross-architecture knowledge distillation in dense retrieval.

Our work is in line with the distillation methods, where LEAD can be a better alternative to the widely used response-based and feature-based methods.

\subsection{Knowledge Distillation}

Knowledge distillation~\cite{DBLP:journals/corr/HintonVD15} aims to transfer knowledge from an effective teacher model to an efficient student model, which can be divided into two categories: response-based methods, and feature-based methods.
The main idea of response-based methods~\cite{lin2022prod, DBLP:conf/sigir/ZengZV22} is to let student model mimic the final predictions of the teacher model~\cite{gou2021knowledge}, which is simple yet effective in different tasks.
In the context of dense retrieval, researchers commonly employ powerful models such as cross-encoder and ColBERT as the teacher model, and leverage the teacher's probability distribution over passages as the supervision signal.
For example, \cite{lin2022prod,DBLP:conf/sigir/ZengZV22,DBLP:journals/corr/abs-2205-09153} incorporate knowledge distillation alongside curriculum strategies to progressively enhance the performance of the student model.
\cite{DBLP:conf/emnlp/RenQLZSWWW21,DBLP:conf/iclr/ZhangGS0DC22} propose on-the-fly distillation, which enables the simultaneous optimization of both the teacher and student models.
However, these methods fail to take into account the intermediate signals of the teacher model, which turn out to be important in very deep neural networks~\cite{DBLP:journals/corr/RomeroBKCGB14}.

Feature-based methods are thought to be the extension of traditional response-based methods and are especially useful in thinner and deeper networks~\cite{gou2021knowledge}, which focus on aligning the internal representations or attention scores of the teacher and the student.
\cite{DBLP:journals/corr/RomeroBKCGB14,xu2020feature,wang2020exclusivity,sun2019patient,haidar2021rail} directly match the internal representations of the teacher and the student, which is not suitable for cross-architecture distillation.
Besides, \cite{DBLP:conf/iclr/ZagoruykoK17, passban2021alp, huang2017like,li2021virt} directly enclose the attention maps of the teacher and the student, which are constrained on used vocabularies and tokenizers.
Therefore, the constraints imposed by architecture, vocabularies, and tokenizers greatly limit the applicability of feature-based methods.

Unlike distillation works on multiple teachers \cite{DBLP:journals/corr/abs-2205-09153,lin2022prod}, LEAD focuses on single-teacher distillation, taking advantage of both response-based and feature-based methods, enclosing the label distributions of intermediate layers, thus leaving no constraints on the vocabularies, tokenizers, or model architectures.

\section{Method}

\begin{figure*}
\centering
\includegraphics[page=1,width=0.8\textwidth]{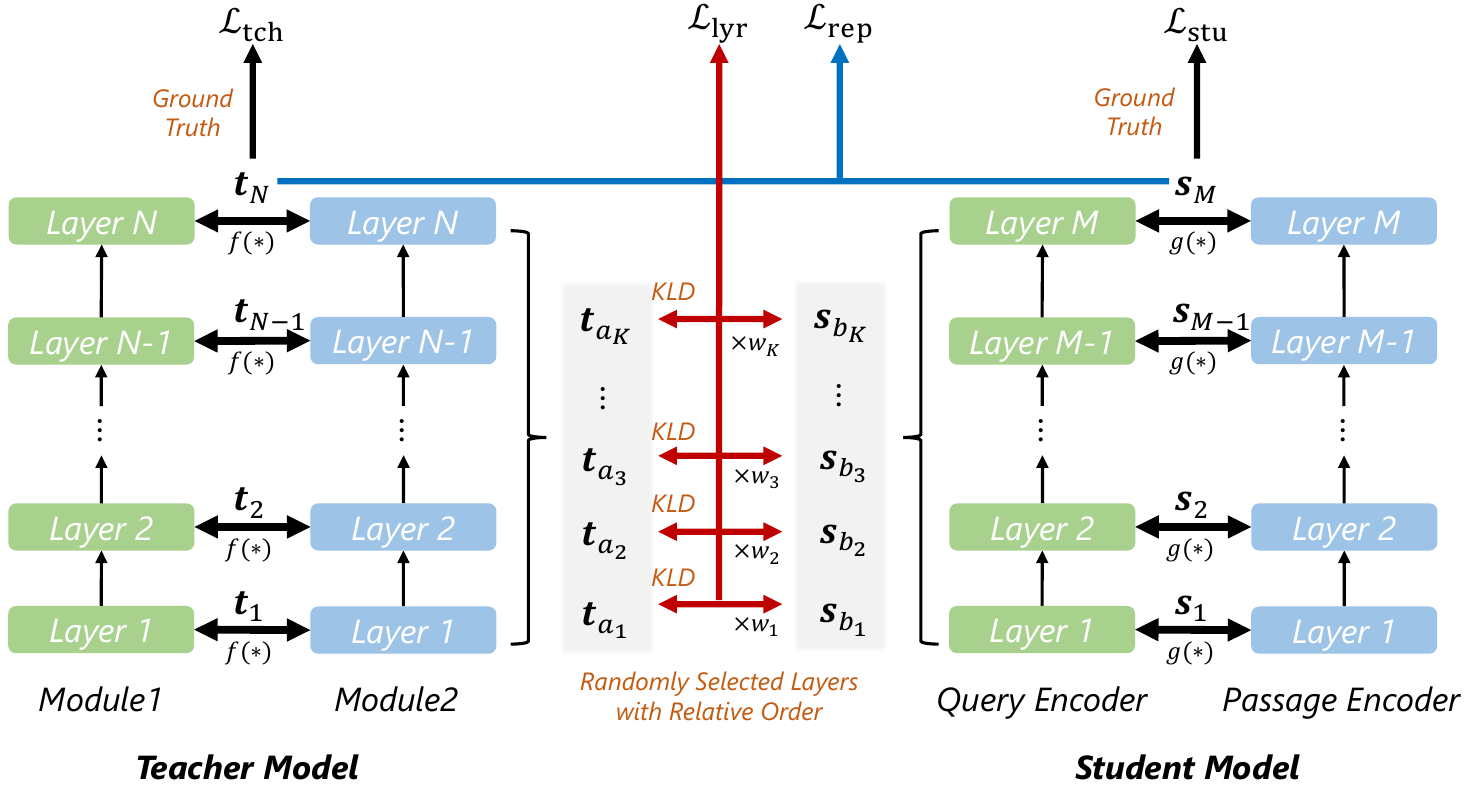}
\caption{
The unified framework of LEAD, which supports various teacher architectures.
Assume the teacher and student model has $N$ and $M$ layers ($N \geq M$), respectively.
The \textit{Module1} and \textit{Module2} correspond to the $E_1$ and $E_2$ of the teacher model, respectively.
$\vt_i$ and $\vs_i$ are the layer features.
$a_1, a_2, ..., a_K$ is the current random selection of teacher layers for layer-wise distillation.
Similarly, $b_1, b_2, ..., b_K$ is for student layers.
The selected layers preserve the original relative order $1 \leq a_1 < a_2 < ... < a_K \leq N$ and $1 \leq b_1 < b_2 < ... < b_K \leq M$.}
\label{fig:framework}
\end{figure*}

\subsection{Task Description}

The target of dense retrieval is to retrieve passages based on learned distributed representations of queries and passages.
Formally, given a query set $\sQ=\{ \vq_1, \vq_2, \dots, \vq_{n} \}$ containing $n$ queries and a collection of passages $\sP=\{ \vp_1, \vp_2, \dots, \vp_{m} \}$ containing $m$ passages, dense retrieval aims to identify the most relevant passages within the corpus $\sP$ for each query $\vq_i$.

\subsection{Unified View of Retrieval Models}
\label{sec:view}

In the research line of dense retrieval, there are several popular models consisting of stacked neural networks, such as dual encoder \cite{DBLP:conf/emnlp/KarpukhinOMLWEC20}, ColBERT \cite{DBLP:conf/sigir/KhattabZ20} and cross encoder \cite{DBLP:conf/naacl/QuDLLRZDWW21}.
Although these models have diverse architectures, they shall have common characteristics.
Therefore, we offer a unified view of these models in this section.

Consider dividing the above-mentioned models into two parallel aligned modules, $E_1$ and $E_2$, which are two piles of neural layers (including Transformer layers and linear layers) with the same layer number.
Given a model with $N$ layers, let's use $E_{1}^{i}$ and $E_{2}^{i}$ ($1 \leq i \leq N$) to denote the encoding of the first $i$ layers of $E_1$ and $E_2$, respectively.
The two major differences between those models are (1) the instantiations of $E_1$ and $E_2$ and (2) the similarity calculation function $f(\cdot)$.
To better elaborate the unified view, we show the following examples.

\textbf{Dual Encoder (DE)} \cite{DBLP:conf/emnlp/KarpukhinOMLWEC20} is the most widely used dense retrieval architecture, which encodes queries and passages into dense vectors separately, calculating the relevance score through the inner product.
For DE, $E_1$ is the query encoder and $E_2$ is the passage encoder.
Both of them are Transformer encoders.
The similarity calculation function $f_{\textrm{DE}}(\cdot)$ is defined as:
\begin{equation}
f_{\textrm{DE}}(\vq, \vp) = E_{Q}(\vq)^{T} \cdot E_{P}(\vp)
\label{equ:descore}
\end{equation}

\textbf{ColBERT (CB)} \cite{DBLP:conf/sigir/KhattabZ20} can be viewed as a more expressive dual-encoder, which delays the interaction between query and passage after encoding.
The instantiation of $E_1$ and $E_2$ is the same as DE.
But the similarity calculation function $f_{\textrm{CB}}(\cdot)$ is defined as:
\begin{equation}
f_{\textrm{CB}}(\vq, \vp) = \sum_{1 \leq x \leq X} \mathop{\textrm{max}}\limits_{1 \leq y \leq Y} E_{Q}(\vq)_x \cdot E_{P}(\vp)_y
\label{equ:colscore}
\end{equation}
where $X$ and $Y$ denote the length of the query and passage token sequence, respectively.
Please note that, following \cite{DBLP:journals/corr/abs-2010-02666}, we remove the punctuation filter and the last linear layer of the encoders to focus on distillation.

\textbf{Cross Encoder (CE)} \cite{DBLP:conf/naacl/QuDLLRZDWW21} has strong abilities to capture the fine-grained relationships between queries and passages within the Transformer encoding.
Much different from DE and CB, for CE, $E_1$ is the query-passage pair encoder $E_{\textrm{CE}}$ and $E_2$ is the projection layer $\vw$ after the Transformer encoder, which is used in a shared manner.
The similarity calculation function $f_{\textrm{CE}}(\cdot)$ is defined as:
\begin{equation}
f_{\textrm{CE}}(\vq, \vp) = \vw^T \cdot E_{\textrm{CE}}([\vq;\vp])
\label{equ:descore}
\end{equation}
where $[;]$ is the concatenation operation.

\subsection{Distilling with Liberal Layer Feature}

Previous response-based distillation methods \cite{DBLP:conf/naacl/QuDLLRZDWW21,DBLP:journals/corr/abs-2010-11386,luan2021sparse} mainly distill knowledge from the output logits of the teacher model to the student model.
However, considering that different model layers try to capture the relationships between queries and passages at different granularities, we believe distilling knowledge from intermediate layers may be good for the student.
Thus, based on the unified view, we propose a liberal feature-based distillation method (LEAD) based on layer-wise alignment, which is shown in \cref{fig:framework}.

\subsubsection{Layer Feature Definition}
In general, the backbone of our model is a pile of Transformer encoders, which abstracts the input information from low-level structure to high-level semantics.
For each layer, we calculate the score distribution between the query and passages in each training instance. This score distribution is denoted as the \textbf{layer feature} and can be defined as follows:
\begin{align}
\vt_{i} &= \mathop{\forall}_{\vp \in \sP^+ + \sP^-} \frac{\exp(f(\vq, \vp))}{\sum_{\vp' \in \sP^+ + \sP^-}\exp(f(\vq, \vp'))}
\label{equ:studentscorepd} \\
\vs_{i} &= \mathop{\forall}_{\vp \in \sP^+ + \sP^-} \frac{\exp(g(\vq, \vp))}{\sum_{\vp' \in \sP^+ + \sP^-}\exp(g(\vq, \vp'))}
\label{equ:studentscorepd}
\end{align}
where $\sP^+$, $\sP^-$ is the relevant and negative passage pool of $\vq$, respectively.
$f(*),g(*)$ is the similarity score function of the teacher model and student model, respectively.
$i$ denotes the $i$-th layer.

\subsubsection{Layer Selection}
In each layer-wise alignment, we need to find a teacher layer for each student layer.
Without loss of the generality, we randomly select $K$ layers ($N \geq M \geq K$) from teacher model $A = \{a_1, a_2,..., a_N\} $ and $K$ layers from student model $B = \{b_1, b_2,..., b_M\}$, respectively, keeping their original relative orders.
Then, we distill knowledge from the $i$-th selected teacher layer $a_i$ to the $i$-th selected student layer $b_i$ by forcing the layer features $\vt_{a_i}$ and $\vs_{b_i}$ to be close:
\begin{equation}
\Ls_{\textrm{lyr}}^{i}=\KL(\vt_{a_i}/\tau_d, \vs_{b_i}/\tau_d)
\label{equ:ctdsoftloss}
\end{equation}
where $\tau_d$ is the distillation temperature.

\subsubsection{Layer Re-weighting}
Although different teacher layers contain different information, some layers may contain noisy information which may be harmful to the model performance.
Thus, we propose a layer re-weighting technique by assigning different weights to involved teacher layers.
The weights are calculated based on the teacher layer features and the ground truth label distribution:
\begin{equation}
w_i = \frac{\exp(-\KL(\vy, \vt_{a_i}) / \tau_l)}{\sum_{\vj = 1}^K \exp(-\KL(\vy, \vt_{a_j}) / \tau_l)}
\label{layer_selection}
\end{equation}
where $\vy$ is the ground truth one-hot label and $\tau_l$ is the layer selection temperature.

At last, the layer distillation loss is the weighted sum of the loss items of all the selected layers.
\begin{equation}
\Ls_{\textrm{lyr}} = \sum_{i=1}^K w_i \times \Ls_{\textrm{lyr}}^{i}
\label{equ:layer_sum}
\end{equation}

\subsection{Joint Training of Teacher and Student}

Previous distillation methods for dense retrieval \cite{DBLP:journals/corr/abs-2010-02666, lin2020distilling, lin2022prod, DBLP:conf/sigir/ZengZV22} usually tune the student with a fixed teacher, while the supervision signals come from the differences in the output label distributions.
We think that the pre-learned teacher models that are converged on the training data could further benefit from distillation.
Therefore, we jointly train the teacher model and the student model, allowing them to learn from each other.

First, we adopt the bi-directional response-based loss as:
\begin{equation}
\Ls_{\textrm{rep}}=\KL(\vt_{N}/\tau_d, \vs_{M}/\tau_d) + \KL(\vs_{M}/\tau_d, \vt_{N}/\tau_d)
\label{equ:ctdsoftloss}
\end{equation}

Second, we calculate the hard loss items by the output distributions and ground truth as:
\begin{align}
\Ls_{\textrm{tch}} &= -\sum_{\vp^+ \in \sP^+} \log\frac{\exp(f(\vq, \vp^+))}{\sum_{\vp \in \sP^+ + \sP^-} \exp(f(\vq, \vp))}
\label{equ:dtdhardloss} \\
\Ls_{\textrm{stu}} &= -\sum_{\vp^+ \in \sP^+} \log\frac{\exp(g(\vq, \vp^+))}{\sum_{\vp \in \sP^+ + \sP^-} \exp(g(\vq, \vp))}
\label{equ:dtdhardloss}
\end{align}

As last, the total loss is the combination of the layer distillation loss, the response-based loss and the hard loss:
\begin{equation}
\Ls_{\textrm{total}} = \Ls_{\textrm{lyr}} + \Ls_{\textrm{rep}} + \Ls_{\textrm{tch}} + \Ls_{\textrm{stu}}
\end{equation}

\subsection{Discussion}

\subsubsection{Extendibility}
Since LEAD is based on the unified view of retrieval models presented in \cref{sec:view}, in which the models are abstracted as piles of neural layers, we think LEAD is extendable to different types of layers.
For example, LEAD also supports variants of DE and CB models, such as appending linear layers to the naive output of DE and CB, reducing the output dimensionalities, which supports building a smaller vector-based index for passages.

\subsubsection{Portability}
As an alternative to traditional response-based distillation methods, LEAD offers a viable option for enhancing dense retrieval models.
It can be seamlessly integrated into existing dense retrieval pipelines, such as curriculum learning \cite{DBLP:conf/sigir/ZengZV22}, data augmentation~\cite{DBLP:conf/emnlp/RenQLZSWWW21,bonifacio2022inpars} and learning with multiple teachers \cite{lin2022prod}, to achieve further performance improvement.
\section{Experiments}
\begin{table*}[ht]
\centering

\caption{The main results against baselines on \textsc{MS-Pas}, \textsc{TREC-Pas-19} and \textsc{TREC-Pas-20}, where ``66M'' and ``110M'' denote 6-layer methods and  12-layer methods, respectively. The superscripts refer to significant improvements compared to TAS-B($^\ast$), SPLADE v2($^\dagger$), ColBERT v2($^\ddagger$), ERNIE-Search($^\S$).}
\vspace{-1em}

\begin{tabular}{l|c|ccc|cc|cc}
\hline
\multirow{2}{*}{\textbf{Method}} &  \multirow{2}{*}{\textbf{\#Params}} & \multicolumn{3}{c|}{\textbf{\textsc{MS-Pas}}} & \multicolumn{2}{c|}{\textbf{\textsc{TREC-Pas-19}}} & \multicolumn{2}{c}{\textbf{\textsc{TREC-Pas-20}}} \\
 & & \textbf{MRR@10} & \textbf{MAP@1k} & \textbf{R@1k} & \textbf{nDCG@10} & \textbf{MAP@1k} & \textbf{nDCG@10} & \textbf{MAP@1k} \\ \hline
BM25 \cite{DBLP:conf/sigir/Yang0L17} &  - & 18.7 & 19.6 & 85.7 & 49.7 & 29.0 & 48.7 &  28.8 \\
DeepCT \cite{DBLP:conf/sigir/DaiC19} & - & 24.3 & 25.0 & 91.0 & 55.0 & 34.1 & 55.6 & 34.3 \\
docT5query \cite{DBLP:journals/corr/abs-1904-08375} & - & 27.2 & 28.1 & 94.7 & 64.2 & 40.3 & 61.9 & 40.7 \\
ANCE \cite{DBLP:conf/iclr/XiongXLTLBAO21} & 110M& 33.0 & 33.6 & 95.9 & 64.8 & 37.1 & 64.6 &  40.8 \\
ADORE \cite{DBLP:conf/sigir/ZhanM0G0M21} & 110M & 34.7 & 35.2 & - & 68.3 & 41.9 & 66.6 &  44.2 \\
ColBERT~\cite{DBLP:conf/sigir/KhattabZ20} & 110M & 36.0 & - & 96.8 & - & - & - & -  \\
Condenser~\cite{Gao2021Condenser} & 110M & 36.6 & - & 97.4 & 69.8 & - & - & -  \\
coCondenser~\cite{gao2021unsupervised} & 110M & 38.2 & - & 98.4 & 71.7 & - & 68.4 & -  \\
\hline
Margin-MSE \cite{DBLP:journals/corr/abs-2010-02666} & 66M & 32.5 & 33.1 & 95.7 & 69.9 & 40.5 & 64.5 & 41.6  \\
TAS-B \cite{DBLP:conf/sigir/HofstatterLYLH21} & 66M & 34.4 & 35.1 & 97.6 & 71.7 & 44.7 & 68.5 & 45.5 \\
SPLADE v2 \cite{DBLP:journals/corr/abs-2109-10086} & 66M & 36.8 & - & 97.9 & \textbf{72.9} & - & - & - \\
CL-DRD \cite{DBLP:conf/sigir/ZengZV22} & 66M & 38.2 & 38.6 & - & 72.5 & 45.3 & 68.7 & 46.5 \\
TCT-ColBERT \cite{DBLP:journals/corr/abs-2010-11386} & 110M & 33.5 & 34.2 & 96.4 & 67.0 & 39.1 & 66.8 & 43.0  \\
RocketQA v2~\cite{DBLP:conf/emnlp/RenQLZSWWW21} & 110M & 38.8 & - & 98.1 & - & - & - & -  \\
AR2~\cite{Zhang2021AR2} & 110M & 39.5 & - & 98.6 & - & - & - & -  \\
ColBERT v2~\cite{DBLP:conf/naacl/SanthanamKSPZ22} & 110M & 39.7 & - & 98.4 & - & - & - & -  \\
ERNIE-Search~\cite{DBLP:journals/corr/abs-2205-09153}  & 110M & 40.1 & - & 98.2 & - & - & - & -  \\
\hline
LEAD (12CB $\rightarrow$ 6DE) & 66M & 37.8\bm{$^{\ast\dagger}$} & 38.3\bm{$^{\ast}$} & 97.4 & 70.4 & 43.0 & 68.9\bm{$^{\ast}$} & 44.1 \\
LEAD (24CB $\rightarrow$ 12DE) & 110M & \textbf{40.4}\bm{$^{\ast\dagger\ddagger\S}$} & \textbf{40.9}\bm{$^{\ast}$} & \textbf{99.0}\bm{$^{\ast\dagger\ddagger\S}$} & 72.5\bm{$^{\ast}$} & \textbf{48.8}\bm{$^{\ast}$} & \textbf{71.2}\bm{$^{\ast}$} & \textbf{48.8}\bm{$^{\ast}$} \\
\hline
\end{tabular}
\vspace{-1em}
\label{tab:main_results_pas1}
\end{table*}

\begin{table*}[ht]
\centering
\caption{The main results against baselines on \textsc{MS-Doc}, \textsc{TREC-Doc-19} and \textsc{TREC-Doc-20},  where ``66M'' and ``110M'' denote 6-layer methods and 12-layer methods, respectively. All baselines are methods without distillation. The superscripts refer to significant improvements compared to ANCE($^\ast$), STAR($^\dagger$), COIL-full($^\ddagger$), ADORE($^\S$).}
\vspace{-1em}

\begin{tabular}{l|c|ccc|cc|cc}
\hline
\multirow{2}{*}{\textbf{Method}} & \multirow{2}{*}{\textbf{\#Params}} & \multicolumn{3}{c|}{\textbf{\textsc{MS-Doc}}} & \multicolumn{2}{c|}{\textbf{\textsc{TREC-Doc-19}}} & \multicolumn{2}{c}{\textbf{\textsc{TREC-Doc-20}}} \\
 & & \textbf{MRR@10} & \textbf{MAP@100} & \textbf{R@100} & \textbf{nDCG@10} & \textbf{R@100}  & \textbf{nDCG@10} & \textbf{R@100} \\ \hline
BM25 \cite{DBLP:conf/sigir/Yang0L17} & - & 27.9 & - & 80.7 & 51.9 & \textbf{39.5} & - & -\\
ME-BERT \cite{luan2021sparse} & 110M & 34.6 & - & - & 61.0 & - & - & - \\
ANCE \cite{DBLP:conf/iclr/XiongXLTLBAO21} & 110M & 37.7  & - & 89.4 & 61.0 & 27.3 & - & -\\
STAR \cite{DBLP:conf/sigir/ZhanM0G0M21} & 110M & 39.0 & - & 91.3 & 60.5 & 31.3 & - & -\\
COIL-full~\cite{gao2021coil} & 110M & 39.7 & - & - & 63.6 & - & - & -\\ 
ADORE \cite{DBLP:conf/sigir/ZhanM0G0M21} & 110M & 40.5 & - & 91.9 & 62.8 & 31.7 & - & -\\
\hline
LEAD (12CB $\rightarrow$ 6DE) & 66M & 40.0\bm{$^{\ast\dagger\ddagger}$} & 40.9 & 88.6 & 63.8\bm{$^{\ast\dagger\ddagger\S}$} & 28.9\bm{$^{\ast}$} & 60.1 & 46.2 \\
LEAD (24CB $\rightarrow$ 12DE) & 110M & \textbf{41.3}\bm{$^{\ast\dagger\ddagger\S}$} & \textbf{42.1} & \textbf{91.9}\bm{$^{\ast\dagger}$} & \textbf{64.0}\bm{$^{\ast\dagger\ddagger\S}$} & 32.5\bm{$^{\ast\dagger\S}$} & \textbf{62.1} &  \textbf{51.4} \\
\hline
\end{tabular}
\label{tab:main_results_doc1}
\end{table*}

\begin{table*}[ht]
\centering
\caption{The main results against different types of teacher models using different distillation methods.
For simplicity, we use ``$n$DE'', ``$n$CB'' and ``$n$CE'' to denote the $n$-layer model.
We also use ``A $\rightarrow$ B'' to denote the distillation process of using A and B as the teacher and student model, respectively.}
\vspace{-1em}
\begin{tabular}{l|l|ccc|cc|cc}
\hline
\multirow{2}{*}{\textbf{Setting}} &
\multirow{2}{*}{\textbf{Method}} & 
\multicolumn{3}{c|}{\textbf{\textsc{MS-Pas}}} & 
\multicolumn{2}{c|}{\textbf{\textsc{TREC-Pas-19}}} & 
\multicolumn{2}{c}{\textbf{\textsc{TREC-Pas-20}}} \\
 & & \textbf{MRR@10} & \textbf{MAP@1k} & \textbf{R@1k} & \textbf{nDCG@10} & \textbf{MAP@1k} &  \textbf{nDCG@10} & \textbf{MAP@1k} \\ \hline
6DE & Student & 32.62 & 33.30  & 96.45 &64.76 & 41.35 & 64.38 & 39.47\\
\hline
\multirow{4}{*}{12DE $\rightarrow$ 6DE} & Teacher & 36.45 & 37.04 & 98.31 & 67.93 & 45.18 & 67.54 & 44.28\\
 & FD & 33.50 &34.10 & 95.69 & 63.97 & 38.93 & 63.75 & 38.21 \\
 & RD & 35.36 & 35.99 & 96.99 & 69.04& 42.05 & 66.31 & 40.85 \\
 & LEAD & 36.26 & 36.85 & 96.51 & 68.91 & 42.11 & 66.26 & 41.49 \\
\hline
\multirow{4}{*}{12CB $\rightarrow$ 6DE} & Teacher & 39.46 & 40.02 & 98.31 & 72.03 & 48.24 & 72.82 & 51.52\\ 
 & FD & 33.92 & 34.56 & 96.12  & 63.69 & 38.93 & 65.54 & 39.71 \\
 & RD & 36.02 & 36.62 & 97.18 & 67.88 & 42.57 & 67.06 & 42.19 \\
 & LEAD & \textbf{37.80} & \textbf{38.32} & \textbf{97.42} & \textbf{70.44} & \textbf{43.01}  & \textbf{68.92} & \textbf{44.09} \\
\hline
\multirow{4}{*}{12CE $\rightarrow$ 6DE} & Teacher & 40.54 & 41.10 & 98.31 & 72.58 & 49.78 & 72.96 & 51.00 \\
 & FD & 34.18 & 34.81 & 95.52 & 65.85 & 40.78 & 65.43 & 40.00 \\
 & RD & 35.74 & 36.31 & 96.55 & 67.98 & 40.34 & 64.68 & 39.89 \\
 & LEAD & 36.73 & 37.23 & 96.13 & 69.12 & 41.05 & 66.63 & 40.12 \\
\hline
\hline
\multirow{2}{*}{\textbf{Setting}} &
\multirow{2}{*}{\textbf{Method}} & 
\multicolumn{3}{c|}{\textbf{\textsc{MS-Doc}}} & 
\multicolumn{2}{c|}{\textbf{\textsc{TREC-Doc-19}}} & 
\multicolumn{2}{c}{\textbf{\textsc{TREC-Doc-20}}} \\
 & & \textbf{MRR@10} & \textbf{MAP@100} & \textbf{R@100} & \textbf{nDCG@10} & \textbf{R@100} &  \textbf{nDCG@10} & \textbf{R@100} \\ \hline
6DE & Student & 36.39 & 37.36 & 86.33 & 61.44 & 27.71 & 56.66 & 46.26 \\
\hline
\multirow{4}{*}{12DE $\rightarrow$ 6DE} & Teacher & 39.88 & 40.79 & 90.31 & 62.36 & 30.73 & 61.51 & 52.75 \\
 & FD & 35.84 & 36.78 & 84.38 & 59.93 & 26.33 & 55.51 & 43.39 \\
 & RD & 38.70 & 39.63 & 87.92 & 63.02 & 28.31 & 57.00 & 42.87 \\
 & LEAD & 39.68 & 40.61 & 88.00 & 63.76 & 27.57 & 59.33 & 44.10 \\
\hline
\multirow{4}{*}{12CB $\rightarrow$ 6DE} & Teacher & 41.58 & 42.44 & 90.31  & 65.69 & 30.73 & 62.04 & 52.75\\ 
 & FD & 36.01 & 36.95 & 84.94 & 59.12 & 26.20 & 56.92 & 43.08 \\
 & RD & 38.45 & 39.37 & 87.98 & 62.11 & 27.88 & 59.09 & 45.70 \\
 & LEAD & \textbf{40.00} & \textbf{40.93} & \textbf{88.63} & \textbf{63.79} & \textbf{28.90} & \textbf{60.05} & \textbf{46.20} \\
\hline
\multirow{4}{*}{12CE $\rightarrow$ 6DE} & Teacher & 41.68 & 42.55 & 90.31 & 61.36 & 30.73 & 58.73 & 52.75 \\
 & FD & 36.42 & 37.32 & 84.25 & 59.26 & 25.92 & 53.82 & 39.66 \\
 & RD & 36.78 & 37.59 & 81.29 & 59.12 & 26.24 & 54.00 & 37.47 \\
 & LEAD & 37.35 & 38.24 & 84.77 & 60.25 & 28.69 & 54.82 & 39.93 \\
\hline
\end{tabular}

\label{tab:main_results2}
\end{table*}

\subsection{Datasets}

Experiments are conducted on several popular retrieval datasets: MS MACRO Passage Ranking (\textsc{MS-Pas}) \cite{DBLP:conf/nips/NguyenRSGTMD16}, TREC 2019 DL Track (\textsc{TREC-Pas-19} and \textsc{TREC-Doc-19}) \cite{DBLP:journals/corr/abs-2003-07820}, MS MARCO Document Ranking (\textsc{MS-Doc}) \cite{DBLP:conf/nips/NguyenRSGTMD16} and TREC 2020 DL Track (\textsc{TREC-Pas-20} and \textsc{TREC-Doc-20}) \cite{DBLP:journals/corr/abs-2102-07662}.

\vspace{-1em}
\subsection{Setting and Implementation}

\subsubsection{Metrics.}
Following previous works~\cite{DBLP:conf/sigir/ZhanM0G0M21,lin2022prod}, for \textsc{MS-Pas}, we report MRR@10, MAP@1k and R@1k on the dev set.
For \textsc{TREC-Pas-19} and \textsc{TREC-Pas-20}, we report nDCG@10 and MAP@1k. 
For \textsc{MS-Doc}, we report MRR@10, MAP@100 and R@100 on the dev set.
For \textsc{TREC-Doc-19} and \textsc{TREC-Doc-20}, we report nDCG@10 and R@100.
We conduct significant tests based on the paired t-test with $p\leq0.01$.

\vspace{-1em}

Following previous work \cite{DBLP:conf/iclr/XiongXLTLBAO21,DBLP:conf/naacl/QuDLLRZDWW21}, we use DE as the student and compute the metrics with \texttt{IndexFlatIP} in faiss\footnote{\url{https://faiss.ai/}}.
While for CB and CE teacher models, we follow \citet{DBLP:journals/corr/abs-2010-02666,lin2020distilling,lin2022prod} and rerank the top-1000 results and the top-100 results retrieved by the 12-layer DE teacher model for \textsc{Pas} (\textsc{MS-Pas}, \textsc{TREC-Pas-19} and \textsc{TREC-Pas-20}) and \textsc{Doc} (\textsc{MS-Doc}, \textsc{TREC-Doc-19} and \textsc{TREC-Doc-20}), respectively.
Please note that when using DE and CB as teacher, cross-batch negatives \cite{DBLP:conf/naacl/QuDLLRZDWW21} are applied during training.

\subsubsection{Training Process.}
First, we use random or BM25 negatives to train a 12-layer DE and get top-100 hard negatives.
Then, we train the teacher and student with the mined hard negatives as warming up.
Finally, we load the trained checkpoints for the teacher model and student model, conducting the knowledge distillation.
The experiments are conducted on 8 NVIDIA V100 GPUs. 
We optimized our models by AdamW~\cite{loshchilov2017decoupled}.

\subsection{Main Results}
To make fair comparisons with existing methods, we conduct experiments using two popular settings: from a 24-layer teacher to a 12-layer DE and from a 12-layer teacher to a 6-layer DE.
Additionally, we use the latter setting to study the properties of our method in subsequent experiments.

\subsubsection{Comparing with Existing Methods}
We compare LEAD with other text retrieval methods on \textsc{MS-Pas}, \textsc{TREC-Pas-19}, \textsc{TREC-Pas-20}, \textsc{MS-Doc}, \textsc{TREC-Doc-19} and \textsc{TREC-Doc-20}.
According to our preliminary study, utilizing CB as the teacher yields better performance, so CB is used as the teacher in this experiment.
We use ``LEAD (12CB $\rightarrow$ 6DE)'' and  ``LEAD (24CB $\rightarrow$ 12DE)'' to denote using 12-layer CB and 24-layer CB as teacher model, respectively.
The results are shown in \cref{tab:main_results_pas1} and \cref{tab:main_results_doc1}.

From the results, we can find that: (1) LEAD (24CB $\rightarrow$ 12DE) demonstrates superior performance on all datasets, and LEAD (12CB $\rightarrow$ 6DE) achieves comparable performance with CL-DRD \cite{DBLP:conf/sigir/ZengZV22}, which is the SOTA 6-layer baseline with the help of curriculum learning.
(2) Comparing methods without KD, LEAD (24CB $\rightarrow$ 12DE) achieves the best performance in terms of nearly all metrics on all datasets.
(3) Among all the baselines with KD,
CL-DRD \cite{DBLP:conf/sigir/ZengZV22} uses DE as teacher model, 
TCT-ColBERT \cite{DBLP:journals/corr/abs-2010-11386} uses CB as teacher model, 
Margin-MSE \cite{DBLP:journals/corr/abs-2010-02666}, SPLADE v2 \cite{DBLP:journals/corr/abs-2109-10086}, RocketQA v1 \cite{DBLP:conf/naacl/QuDLLRZDWW21}, RocketQA v2~\cite{DBLP:conf/emnlp/RenQLZSWWW21}, AR2~\cite{Zhang2021AR2} and ColBERT v2~\cite{DBLP:conf/naacl/SanthanamKSPZ22} use CE as teacher model, 
and TAS-B \cite{DBLP:conf/sigir/HofstatterLYLH21} and ERNIE-Search~\cite{DBLP:journals/corr/abs-2205-09153} use CB and CE as teachers.
LEAD (24CB $\rightarrow$ 12DE), which uses CB as teacher model, demonstrates the best performance on nearly all the metrics.
The reason why SPLADE v2 \cite{DBLP:journals/corr/abs-2109-10086} performs better on nDCG@10 of \textsc{TREC-Pas-19} might be that the max pooling operation reduces noises from unrelated tokens, resulting in better query and passage embeddings.

\subsubsection{Comparing with Different KD Strategies}
To further compare with other knowledge distillation methods in a fair setting, we compare with different knowledge distillation strategies\footnote{For RD, we directly enclose the output distribution of teacher and student. For FD, like \cite{DBLP:journals/corr/abs-2112-04195}, we enclose the internal attention map of teacher and student.} like response-based methods (\textbf{RD}) \cite{DBLP:conf/naacl/QuDLLRZDWW21} and feature-based methods (\textbf{FD}) \cite{DBLP:journals/corr/abs-2112-04195}.
Concretely, we use \textbf{12-layer DE}, \textbf{12-layer CB} and \textbf{12-layer CE} as the teacher model respectively, and conduct knowledge distillation with a 6-layer DE student.
The results are illustrated in \cref{tab:main_results2}. 

As we can see from the table, it is obvious that we can get the following conclusions:
(1) Compared with the original 6-layer DE student, LEAD can achieve further improvements with various teacher models, especially for MRR@10, MAP@100, MAP@1k and nDCG@10.
(2) When training with the same teacher model, generally speaking, RD is better than FD in all the comparisons, indicating that FD is not easy to work on distillation for dense retrieval.
Furthermore, LEAD outperforms other KD strategies, such as FD and RD, on nearly all the datasets and metrics, showing the effectiveness of our method.
(3) Among the explored teacher models, 12-layer CB is the best for boosting a 6-layer DE student when using LEAD.
We guess the reason may be twofold.
First, CB and DE share a similar architecture while the late interaction brings CB better performances, which further leads to better distillation results.
Second, although the CE teacher outperforms the CB teacher, training with CB in LEAD can benefit from the cross-batch negatives \cite{DBLP:conf/naacl/QuDLLRZDWW21}, which is limited in CE.
Therefore, we use 12-layer CB as the teacher model for LEAD in the rest experiments.

\subsection{Effect on Layer Selection}
\label{sec:layer_selection}
In this section, we explore the effects of two major factors in the layer selection of LEAD.
\begin{table}[t]
\centering

\caption{The results of different layer number $K$ on \textsc{MS-Pas} and \textsc{MS-Doc}.}
\vspace{-1em}
\begin{tabular}{c|cc|cc}
\hline
\multirow{2}{*}{\textbf{Setting}} & \multicolumn{2}{c|}{\textbf{\textsc{MS-Pas}}}        & \multicolumn{2}{c}{\textbf{\textsc{MS-Doc}}} \\
                       & \textbf{MRR@10} & \multicolumn{1}{c|}{\textbf{MAP@1k}} & \textbf{MRR@10}        & \textbf{MAP@100}        \\ \hline
$K=1$                      & 37.04  & 37.62 & 39.46 & 40.37 \\
$K=2$                      & 37.53  & 38.05 & 39.73 & 40.65 \\  
$K=3$                      & 37.59  & 38.06 & 39.63 & 40.57\\ 
$K=4$                      & 37.66  & 38.17 & 39.90 &  40.80 \\ 
$K=5$                      & \textbf{37.80} & \textbf{38.32} & 39.98 &  40.88 \\ 
$K=6$                      & 37.68  & 38.13 & \textbf{40.00} & \textbf{40.93} \\ \hline
\end{tabular}
\label{tab:layer_selection1}
\end{table}

\begin{table}[t]
\centering
\caption{The result of different layer selection strategies on \textsc{MS-Pas}. The superscripts refer to significant improvements compared to Weighted($^\ast$), Last($^\dagger$), Skip($^\ddagger$).}
\vspace{-1em}
\begin{tabular}{l|ccc}
\hline
\textbf{Method} & \textbf{MRR@10} & \textbf{MAP@1k} & \textbf{R@1k}  \\ \hline
Random & \textbf{37.80}\bm{$^{\ast\dagger\ddagger}$} & \textbf{38.32}\bm{$^{\ast\dagger\ddagger}$} & \textbf{97.42}\bm{$^{\dagger\ddagger}$} \\
Weighted & 37.13 & 37.67 & 97.38\\
Last & 37.05 & 37.61 & 97.08\\
Skip & 37.18 & 37.73 & 97.05\\
\hline
\end{tabular}
\vspace{-1em}
\label{tab:layer_selection2}
\end{table}

\subsubsection{Study on Layer Number $K$}
We tune the distillation layer number $K$ in the range of $\{1, 2, 3, 4, 5, 6\}$ on \textsc{MS-Pas} and \textsc{MS-Doc}, and observe the trends of student performances, which is shown in \cref{tab:layer_selection1}.
We can observe that the model performance gradually improves with the increase of $K$.
However, the model performance reaches the peak at $K=5$ on \textsc{MS-Pas} and $K=6$ on \textsc{MS-Doc}, respectively.
This is because, with more distillation layers, the student can absorb more knowledge from the teacher, which benefits model performance, but eventually the best layer number highly depends on the dataset characteristic.
Therefore, we use $K=5$ for \textsc{MS-Pas} and $K=6$ for \textsc{MS-Doc} in subsequent experiments.

\subsubsection{Study on Selection Strategies}
We further investigate the performance of different layer selection strategies including Random, Weighted, Last and Skip in \cite{sun2019patient,haidar2021rail}. \textbf{Random} refers to randomly selecting layers for teacher model and student model, which is used in LEAD. \textbf{Weighted} refers to randomly selecting layers for student model while selecting layers for teacher model according to \cref{layer_selection}, where a higher weight indicates a higher probability of being chosen. \textbf{Last} refers to distilling information from the last layers. And \textbf{Skip} refers to distilling information across every $k=2$ layers.
It is worth mentioning that the layer selection choice keeps dynamically changing during training in Random while remaining fixed in Last and Skip or changing a little in Weighted. 
Specifically, for Last strategy, the layer choice for student model is \{2, 3, 4, 5, 6\} and the layer choice for teacher model is \{8, 9, 10, 11, 12\}.
For Skip strategy, we set $k=2$ , the layer choice for student model is \{2, 3, 4, 5, 6\} and the layer choice for teacher model is  \{1, 3, 5, 7, 9\}.
The result of the selection strategies on \textsc{MS-Pas} is shown in \cref{tab:layer_selection2}. 

From the result, we can find that our layer selection strategy Random is better than Weighted, Last and Skip. Presumably, it might be because they only distill information from part of intermediate layers, which may lose information from unchosen layers. However, by distilling information from dynamically changing intermediate layers, Random captures more diverse distributions of richer semantics from low-level to high-level.

\subsection{Ablation Study}
\begin{table}[t]
\centering

\caption{Ablation study results on \textsc{MS-Pas}.}
\vspace{-1em}
\begin{tabular}{l|cccc}
\hline
\textbf{Method} & \textbf{MRR@10} & \textbf{MAP@1k} & \textbf{R@1k}  \\ \hline
LEAD & \textbf{37.80} & \textbf{38.32} & \textbf{97.42}   \\
-- w/o Joint Training & 37.24 & 37.79 & 97.12   \\
-- w/o Layer Re-weighting  & 36.46 & 37.06 & 97.08  \\
\hline
RD & 36.02 & 36.62 & 97.18 \\
\hline
\end{tabular}
\vspace{-0.5em}
\label{tab:ablation}
\end{table}
\begin{table}[t]
\centering
\caption{Extendibility study on \textsc{MS-Pas}.}
\vspace{-1em}
\begin{tabular}{l|cccc}
\hline
\textbf{Method}   & \textbf{MRR@10} & \textbf{MAP@1k}    \\ \hline
RD & 36.02 & 36.62  \\
RD + linear & 34.53 & 35.10  \\
\hline
FD & 33.92 & 34.56 \\
FD + linear & 33.20 & 33.84 \\
\hline
LEAD & 37.80 & 38.32  \\
LEAD + linear & 36.32 & 36.94 \\
\hline
\end{tabular}
\vspace{-1em}

\label{tab:extend}
\end{table}

\begin{table*}[h]
\centering

\caption{The results of LEAD (24CB $\rightarrow$ 12DE) using different PLMs, where ``110M'' denotes 12-layer methods.}
\vspace{-1em}
\begin{tabular}{l|c|ccc|cc|cc}
\hline
\multirow{2}{*}{\textbf{Method}} &  \multirow{2}{*}{\textbf{\#Params}} & \multicolumn{3}{c|}{\textbf{\textsc{MS-Pas}}} & \multicolumn{2}{c|}{\textbf{\textsc{TREC-Pas-19}}} & \multicolumn{2}{c}{\textbf{\textsc{TREC-Pas-20}}} \\
 & & \textbf{MRR@10} & \textbf{MAP@1k} & \textbf{R@1k} & \textbf{nDCG@10} & \textbf{MAP@1k} & \textbf{nDCG@10} & \textbf{MAP@1k} \\ \hline
LEAD~(ERNIE-2.0-Base) & 110M & 38.52 & 39.08  & 97.69 &  \textbf{74.13}  & 46.08 &  67.96  & 44.24  \\
LEAD~(coCondenser) & 110M & 38.96 &  39.52  & 98.61 & 71.13 &  45.75 & 68.69 & 45.24 \\
LEAD (MASTER) & 110M & \textbf{40.40} & \textbf{40.90} & \textbf{98.98} & 72.54 & \textbf{48.81} & \textbf{71.23} & \textbf{48.75} \\
\hline
\hline
\multirow{2}{*}{\textbf{Method}} & \multirow{2}{*}{\textbf{\#Params}} & \multicolumn{3}{c|}{\textbf{\textsc{MS-Doc}}} & \multicolumn{2}{c|}{\textbf{\textsc{TREC-Doc-19}}} & \multicolumn{2}{c}{\textbf{\textsc{TREC-Doc-20}}} \\
 & & \textbf{MRR@10} & \textbf{MAP@100} & \textbf{R@100} & \textbf{nDCG@10} & \textbf{R@100}  & \textbf{nDCG@10} & \textbf{R@100} \\ \hline
LEAD (ERNIE-2.0-Base) & 110M & 40.75  &  41.65 & 88.71 & 61.34 & 27.07 & 57.87 & 45.87  \\
LEAD (coCondenser) & 110M & 40.90 & 41.81 & 90.58 & 61.99 & 31.44 & 60.99 & 49.81\\
LEAD (MASTER)& 110M & \textbf{41.25} &  \textbf{42.14} & \textbf{91.88} & \textbf{64.01} & \textbf{32.50} & \textbf{62.08} &  \textbf{51.44} \\
\hline
\end{tabular}
\vspace{-1em}
\label{tab:12layer main_results_doc1}
\end{table*}

To show the effectiveness of each component, we conduct an ablation study on LEAD.
Since the layer selection has been discussed in \cref{sec:layer_selection}, we focus on the joint training and the layer re-weighting, whose results are shown in \cref{tab:ablation}.

First, to completely forbid the joint training, we directly remove the teacher's hard loss item $\Ls_{\textrm{tch}}$ and $\KL(\vs_{M}/\tau_d, \vt_{N}/\tau_d)$ in $\Ls_{\textrm{rep}}$ from the total loss, freezing the teacher model during training.
It is clear that after removing the joint training from LEAD, the student's performances on all the metrics decreased.
The reason may be that the information of the teacher model and the student model can complement each other.

Second, we replace the weight $w_i$ with a constant $1/K$ in \cref{equ:layer_sum} to show the effects of layer re-weighting.
As we can see, removing the layer re-weighting causes a larger performance drop on \textsc{MS-Pas} compared with the joint training, but still outperforms RD on nearly all the metrics.
The reason may be that by assigning higher weights to more informative layers, the noises from uninformative layers can be largely reduced.

\subsection{Further Study}

\subsubsection{Extendibility}

To verify the extendibility of LEAD, we append a linear layer 768 $\rightarrow$ 768 after the student model and teacher model, and observe the performance of RD, FD and LEAD on \textsc{MS-Pas}.

As shown in \cref{tab:extend}, we can find that after appending the linear layer, the performances of all distillation methods drop. However, our method LEAD+linear still achieves the best performance, which demonstrates the extendibility of our method.

\subsubsection{Portability}
\begin{table}[t]
\centering
\caption{Portability study on \textsc{MS-Pas}.}
\vspace{-1em}
\resizebox{\columnwidth}{!}{
\begin{tabular}{l|c|c|cc}
\hline
\textbf{Method} & \textbf{Step} & \textbf{Description} & \textbf{MRR@10} & \textbf{MAP@1k}       \\ 
\hline
\multirow{3}{*}{Teacher} & T1 & 12DE & 36.45 & 37.04 \\
 & T2 & 12CB & 39.46 & 40.02 \\
 & T3 & 12CE & 40.54 & 41.10 \\
\hline
Student & S0 & 6DE  & 32.62 & 33.30 \\
\hline
\multirow{3}{*}{PROD*} & S1 & 12DE $\rightarrow$ 6DE & 35.36 & 36.00 \\
 & S2 & 12CB $\rightarrow$ 6DE & 35.92 & 36.58 \\
 & S3 & 12CE $\rightarrow$ 6DE & 36.30 & 36.92 \\ 
\hline
\multirow{3}{*}{PROD* + LEAD} & S1 & 12DE $\rightarrow$ 6DE & 36.26 & 36.85\\
 & S2 & 12CB $\rightarrow$ 6DE & 37.64 & 38.18\\
 & S3 & 12CE $\rightarrow$ 6DE & 37.82 & 38.34 \\
\hline
\end{tabular}
}
\vspace{-1em}
\label{tab:plug_in}
\end{table}

To verify the portability of LEAD, we follow PROD~\cite{lin2022prod} and choose three teacher models for continual distillation, which include DE, CB and CE.
PROD* refers to using RD in the PROD distillation process, while PROD*+LEAD refers to using LEAD in the PROD distillation process.
The results on \textsc{MS-Pas} are shown in \cref{tab:plug_in}.

From the result, we can find that:
(1) The performance of student model gradually improves in each step for both PROD* and PROD*+LEAD. However, the performance gain of PROD*+LEAD in each step is better than that of PROD*, which leads to better performance of the final student model. The reason may be that compared with RD, LEAD can fully utilize the information from the intermediate layers of the teacher model.
(2) Compared with LEAD which solely uses CB as teacher model, the result of PROD*+LEAD after continual distillation gets little improvement. This may be due to the fact that the multi-teachers used in the continual distillation share similar abilities.
Overall, the experimental results prove the feasibility of plugging LEAD into the existing distillation paradigm and further improving the model performance.

\begin{table}[t]
\centering
\caption{The results of teacher model before LEAD and after LEAD on \textsc{MS-Pas} and \textsc{MS-Doc}.}
\vspace{-1em}
\resizebox{\columnwidth}{!}{
\begin{tabular}{l|l|cc|cc}
\hline
\multirow{2}{*}{\textbf{Setting}} & \multirow{2}{*}{\textbf{Method}} & \multicolumn{2}{c|}{\textbf{\textsc{MS-Pas}}}        & \multicolumn{2}{c}{\textbf{\textsc{MS-Doc}}} \\
 & & \textbf{MRR@10} & \multicolumn{1}{c|}{\textbf{MAP@1k}} & \textbf{MRR@10}        & \textbf{MAP@100}        \\ \hline
\multirow{2}{*}{12DE} & Before & 36.45 & 37.04 & 39.88 & 40.79  \\
 & After & 37.66 & 38.21 & 40.97 & 41.85 \\
\hline
\multirow{2}{*}{12CB} & Before & 39.46 & 40.02 & 41.58 & 42.44 \\
 & After & 40.49 & 41.00 & 41.94 & 42.79 \\
\hline
\multirow{2}{*}{12CE} & Before & 40.54 & 41.10 &  41.68 & 42.55 \\
& After & 41.35 & 41.87 & 43.12 & 43.93 \\
\hline
\end{tabular}
}
\vspace{-1em}
\label{tab:teacher}
\end{table}

\subsubsection{Joint Training Benefits Teachers}

We propose to jointly train the teacher and student with the intuition that they can complement each other.
To verify it, we evaluate the performance of teacher model before and after LEAD on \textsc{MS-Pas} and \textsc{MS-Doc}.

From the result shown in \cref{tab:teacher}, we can find that the performance of teacher models can be greatly improved after LEAD, which verifies that the teacher model can also be improved within the learning procedure. We leave further analysis of this phenomenon as future work.

\subsubsection{Performance Comparison Using Different PLMs}

To verify the generality of LEAD with different PLMs, we further experiment with LEAD using different PLMs. 
Specifically, we use ERNIE-2.0-Large to initialize the 24-layer CB teacher model and use ERNIE-2.0-Base, coCondenser and MASTER to initialize the 12-layer DE student model, respectively.

As shown in Table \ref{tab:12layer main_results_doc1}, the results demonstrate that LEAD achieves consistent superior performance across diverse PLMs, showcasing the compatibility of LEAD with various PLMs.
Besides, we can also find that using MASTER to initialize the 12-layer dual encoder student model yields the best performance, which are reported in the main results.
\section{Conclusion}
In this paper, we propose a novel knowledge distillation method LEAD for dense retrieval, which aligns the layer features of student and teacher, emphasizing more on the informative layers by layer re-weighting technique.
Through extensive experiments across six widely used benchmarks, our results highlight that LEAD consistently outperforms response-based and feature-based methods by a significant margin.
Moreover, LEAD is extendable to more linear layers and is easy to plug into the existing distillation pipelines.
This study confines its focus to the conventional single-teacher knowledge distillation approach, omitting the exploration of alternative techniques such as curriculum learning and data augmentation.
As for future work, we hope to investigate whether these techniques can be integrated into LEAD for further improvement.
\section{Ethical Considerations }
This paper presents an innovative distillation approach for dense retrieval, which can enhance web search system performance. It can also provide valuable insights to fellow researchers in this field, potentially yielding positive societal impacts. We do not foresee any form of negative societal impact induced by our work.

\bibliographystyle{ACM-Reference-Format}
\balance
\bibliography{LEAD}

\end{document}